\documentclass[showpacs,preprintnumbers,superscriptaddress,amsmath,reprint,aps,pra,twocolumn]{revtex4-1}

\usepackage{bm}
\usepackage{graphicx} 
\usepackage{color}
\usepackage{amsmath}
\usepackage{booktabs}

\newcommand{\EEA}{\end{eqnarray}}
\newcommand{\EEAN}{\end{eqnarray*}}

\begin{document}
 
\title{Vibrational quenching and reactive processes of weakly bound molecular ion-atom collisions at cold temperatures}

\author{Jes\'{u}s P\'{e}rez-R\'{i}os}

\affiliation{School of Natural Sciences and Technology, Universidad del Turabo, Gurabo, PR00778, USA }

\date{\today}

\begin{abstract}

We present a study of vibrational quenching and chemical processes of molecular ions immersed in an ultracold atomic gas by means of the quasi-classical trajectory (QCT) method. In particular, BaRb$^+(v)$ + Rb collisions are studied at cold temperatures revealing a highly efficient energy transfer between the translational degrees of freedom and internal degrees of the molecular ion. Thus leading to a large vibrational quenching cross section which follows the Langevin capture model prediction. These results play a key role on the understanding of relaxation phenomena of cold molecular ions in ultracold gases as well as in the fate of the molecules emerging trough ion-neutral-neutral three-body recombination in an cold and dense environment.


\end{abstract}

\maketitle

\section{Introduction}

Cold hybrid ion-neutral systems present a unique playground for exploring novel cold chemistry scenarios~\cite{Tomza2017,Willitsch2008,Harter2014}, the design of new high precision spectroscopy techniques~\cite{Brunken2017}, the development of effective quantum logic spectroscopy approaches~\cite{Mur2012,Leibfried2012,Wolf2016}, novel quantum information protocols~\cite{Monroe2016QI,Doerk2010,Monroe2010} and the simulation of complex many-body Hamiltonians~\cite{Monroe2017,Monroe2015,Blatt2012}. In cold chemistry, the major driving force is the study of molecular ions colliding with neutrals. This could elucidate the ultimate nature of ion-neutral collisions including stereochemical effects, the possibility of sympathetic cooling of molecular ions with neutrals, and the design of novel spectroscopy techniques for rotational states of molecular ions~\cite{Schlemmer1999,Schlemmer2002,Brunken2017}. All or these applications rely on the control of the internal degrees of freedom of the molecular ion. However, molecular ion-neutral collisions lead to a decoherence process that depends on the collision energy, in which the collision energy is effectively transfer to the internal degrees of freedom of the molecular ion (rotation and vibration) and vice versa. 

 This process, known as relaxation, has largely been studied in chemical physics~\cite{Bird,McCourt,Zhdanov,Montero2014}. It has mainly been explored for neutral species at room temperatures, but recently some efforts have been devoted to the study of rotational and vibrational relaxation or quenching for molecular ion-neutral collisions at cold temperatures~\cite{Hudson2016,Gianturco2011,Wester2015,Stoecklin2005,Stoecklin2008,Stoecklin2011,JPR2016}. These studies are focussed on vibrational and/or rotational relaxation of deep vibrational states, since it is the most usual experimental condition. Nevertheless, recently it has been shown that three-body recombination of ions in a high dense ultracold neutral media leads to an efficient production of molecular ions, but in highly vibrational states~\cite{Krukow2016,JPR2015}, which relaxation mechanism are unknown.

 In this paper we present the study of vibrational relaxation and reactive processes of highly vibrational excited molecular ions in a neutral gas at cold temperatures. In particular, we will study BaRb$^+$ - Rb collisions which are relevant after a single Ba$^+$ is brought in contact with a very dense cloud of ultracold Rb atoms and three-body recombination occurs~\cite{Krukow2016}. Our theoretical approach is based on quasi-classical trajectory (QCT) calculations fueled by the satisfactory results of classical trajectory calculations for ion-neutral-neutral three-body recombination~\cite{Krukow2016,RMP}.
 
 The paper is structured as follows: in Sec. II the QCT calculations method for molecular ion-neutral collisions  is introduced. Next, the theoretical approach is applied to BaRb$^+$ - Rb and the results for vibrational quenching and different chemical reactive channels are shown in Sec. III. These results fueled us to study the possibility of sympathetic cooling of molecular ions in highly excited vibrational states as shown in Sec. IV, and finally, in Sec. V a brief summary and conclusions are presented.


\section{Quasi-classical trajectory (QCT) calculations}

Quasi-classical trajectory (QCT) calculations is a well established technique in chemical physics since the pioneering work of Karplus et al. for the study of H$_2$-H vibrational relaxation~\cite{Karplus,Truhlarbook}. In this approach the initial state of the molecule is prepared in a discrete internal energy state emulating the quantal rovibrational states through the  Wentzel, Kramers and Brillouin (WKB) or semi-classical quantization rule. The dynamics of the nuclei in the potential energy surface (PES) follows Newton's laws of classical mechanics. Thus, once the trajectory has begun, the constrain over the initial conditions relaxes leading to the sampling of the given phase-space. In QCT, the same semi-classical quantization rule is applied to the analysis of product states, as shown below. 

\begin{figure}[t]
\centering\includegraphics[width=1.0\columnwidth]{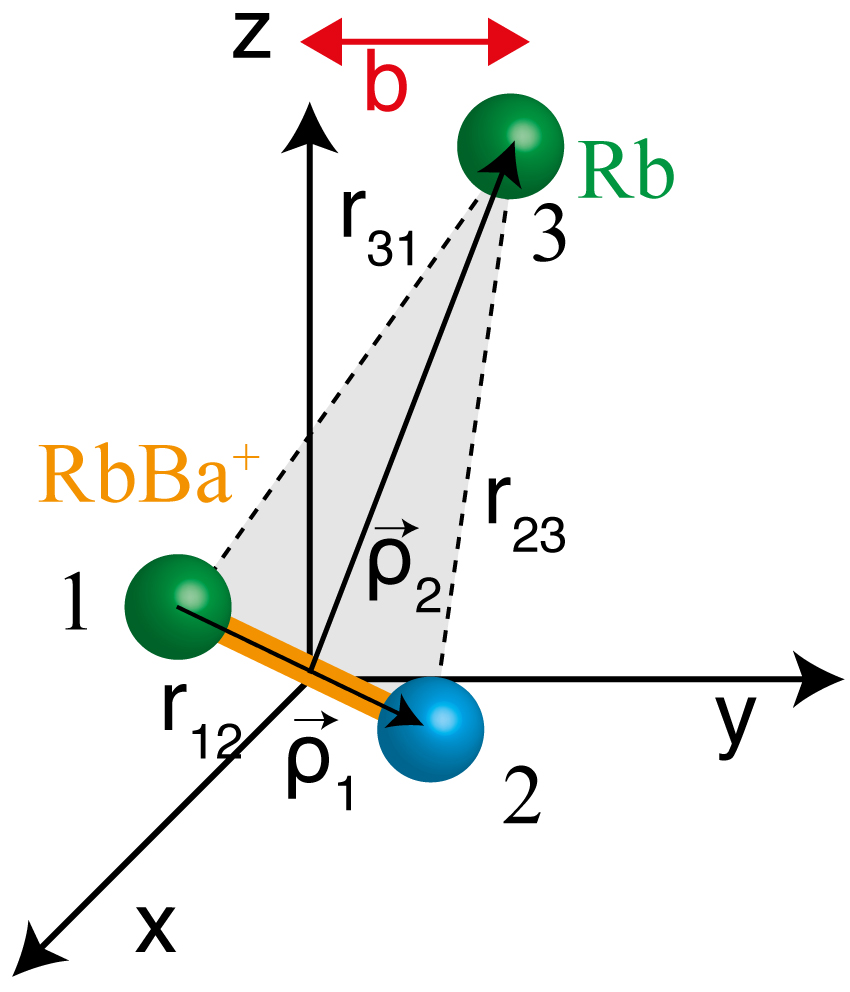}
\caption{Schematic of atom-molecule collision: impact parameter $b$, Jacobi vectors ($\vec{\rho}_1$,$\vec{\rho}_2$) and its relation with the interatomic vectors $\vec{r}_{ij}$, with $(i,j)$=1,2,3 and $i\neq j$  }
\label{fig1}
\end{figure}

The Hamiltonian of three interacting particles under the potential energy surface $V(\vec{r}_1,\vec{r}_2,\vec{r}_3)$ reads as

\begin{equation}
H=\frac{\vec{p}_1^2}{2m_1}+\frac{\vec{p}_2^2}{2m_2}+\frac{\vec{p}_3^2}{2m_3}+V(\vec{r}_1,\vec{r}_2,\vec{r}_3),
\end{equation}

\noindent
where $\vec{r}_i$ and $\vec{p}_i$ stand for the vector position and momentum of the $i$-th atom, respectively. This Hamiltonian is further simplified by using the Jacobi coordinates~\cite{Truhlarbook,JPR2014,RMP} shown in Fig.~\ref{fig1} and neglecting the trivial center of mass motion 

\begin{equation}
H=\frac{\vec{P}_1^2}{2m_{12}}+\frac{\vec{P}_2^2}{2m_{3,12}}+V(\vec{\rho}_1,\vec{\rho}_2),
\end{equation}

\noindent
with $m_{12}=(m_1^{-1}+m_2^{-1})^{-1}$ and $m_{3,12}=(m_3^{-1}+m_{12}^{-1})^{-1}$. Here $\vec{\rho}_1$ represents the Jacobi vector describing the molecule position,  $\vec{P}_1$ stands for its conjugate momentum, whereas $\vec{\rho}_2$ stand for the motion of the atom with respect to the center of mass of the molecule and $\vec{P}_2$ is its momentum. 

The motion of the nuclei is tracked by solving Hamilton's equations of motion, which in this case are

\begin{eqnarray}
\label{Hamilton1}
\frac{d \rho_{i,\alpha}}{dt}=\frac{\partial H}{\partial P_{i,\alpha}} \\
\label{Hamilton2}
\frac{dP_{i,\alpha}}{dt}=-\frac{\partial H}{\partial \rho_{i,\alpha}}, 
\end{eqnarray}

\noindent
where $i=1,2$ and $\alpha$ denotes the different Cartesian components of the Jacobi vectors.

\subsection{Initial conditions}
\label{theory}

Let us assume an atom-molecule collision with a given collision energy $E_k$ in the CM frame, where the atom is placed along the $z$-axis at a given distance $R$ from the CM of the molecule, positioned at the origin of the coordinate system. In this case, after specifying the impact parameter $b$ the vectors $\vec{\rho}_2$ and $\vec{P}_2$ are fully characterized~\cite{Truhlarbook}. On the other hand, $\vec{\rho}_1$ and $\vec{P}_1$ are characterized by the initial rovibrational state $(v,j)$ of the molecule. The size of the molecule $|\rho_1|$ is given by the outer classical turning point $r_{+}$ which is one of the zeros of the molecular kinetic energy 

\begin{equation}
T(r)=E_{\text{int}}-V(r)-\frac{\hbar^2j(j+1)}{2m_{12} r^2},
\end{equation}

\noindent
where $E_{\text{int}}$ represents the rovibrational energy and $V(r)$ stands for the molecular potential energy curve~\footnote{Here $r=r_{12}$ based on Fig~\ref{fig1}}. Here, we chose $P_{1}=\hbar j(j+1)/r_{+}$. Then, the initial state of the molecule is fully described  by specifying the azimuthal angle $\theta$ of the molecule with respect to the $z$-axis, its polar angle in the $x-y$ plane $\phi$ and the angle $\eta$ between the angular momentum of the molecule $\vec{J}=\vec{\rho}_1\times \vec{P}_1$ and a normal vector to the molecular axis. 

Finally, to fully randomize rotation and vibration of the molecule in each collision one needs to choose properly the initial distance between the incoming particle and the target $R=R_0+\frac{\chi P_2\tau_{v,j}}{\mu_{3,12}}$ ($R=|\vec{\rho}_2|$), where $R_0$ is some fixed atom-molecule distance in which the potential energy is negligible in comparison with the collision energy,  $\chi \in[0,1]$ is a uniform random variable and


\begin{equation}
\label{eq}
\tau_{v,j}=\sqrt{2m_{12}}\int_{r_{-}}^{r_{+}} \frac{dr}{\sqrt{2\mu \left[E_{\text{int}}-V(r)-\frac{\hbar^2j(j+1)}{2m_{12} r^2}  \right]}},
\end{equation}

\noindent
is the vibrational period. In Eq.(\ref{eq}), $r_{-}$ represents the inner classical turning point for the given rovibrational energy of the molecule. 

\subsection{Reaction products}
We are interested in two possible product states: a change of the rovibrational state of the molecule, the so-called vibrational relaxation or quenching, and reactive collisions. The first of the processes is fully characterized by the final rovibrational state of the molecule which is determined through the vibrational quantum number

\begin{equation}
\label{nup}
v'=-\frac{1}{2}+\frac{1}{\pi \hbar}\int_{r_{-}}^{r_{+}} \sqrt{2m_{12} \left[E_{int}-V(r)-\frac{\hbar^2j'(j'+1)}{2m_{12} r^2}  \right]}dr,
\end{equation} 

\noindent
where $E_{int}=P_1^2/2m_{12}+V(r)$ represents the internal energy of the molecule, $j'$ is the rotational quantum number given by $ j'=-1/2+\sqrt{\vec{J}'\cdot\vec{J}'/\hbar^2}$ where $\vec{J}'=\vec{\rho}_1 \times \vec{P}_1$ at the final propagation time.

Reactive events are of two kinds: 

\begin{itemize}

\item Dissociation: three free atoms as a final state. This processes becomes operative when the collision energy is larger than the binding energy of the molecule. 


\item Molecular formation: formation of a new product not present in the reactants. This is identified by looking at the internal energy of the atom pairs in the complex and finding which is negative and by applying Eq.~\ref{nup}, but with the right interatomic potential and reduce mass, the vibrational and rotational level of the product state can be addressed. 

\end{itemize}

\begin{figure*}[t!!]
\centering\includegraphics[width=1.0\textwidth]{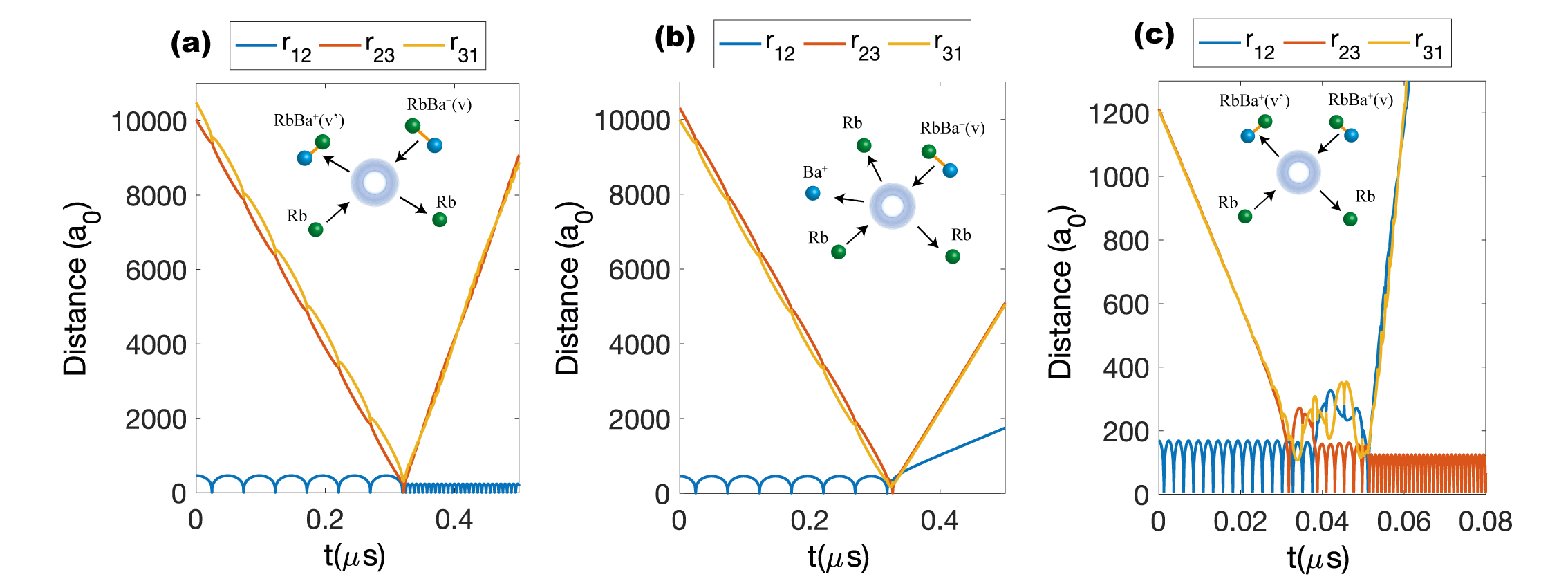}
\caption{Trajectories for RbBa$^+(v)$ + Rb collisions. Panel (a): trajectory for $b=0$a$_0$, $v=195$ and E$_{\text{k}}$=10~mK leading to a quenching process where $v'=191$. Panel (b) a trajectory for  $b=90$~a$_0$, $v=195$ and E$_{\text{k}}$=10mK leading to a dissociation process. Panel (c): a collision for $b=0$a$_0$, $v=187$ and E$_{\text{k}}$=10~mK leading to a quenching collision with $v'=184$. In this collision the quenching is produced through an exchange of the Rb atom forming the initial molecular ion and the colliding Rb atom. The inset of each figure schematically represents the physics process under study. }
\label{fig2}
\end{figure*}

\subsection{Cross section}

Classically the cross section associated with a given process 

\begin{itemize}
\item Quenching ($q$); AB$^+(v)$ +C $\rightarrow$ AB$^+(v')$ + C 
\item Dissociation ($d$); AB$^+(v)$ +C $\rightarrow$ A + B$^+$ + C
\item Reaction ($r$);  AB$^+(v)$ +C $\rightarrow$ CB$^+(v'')$ + A;  \\ AB$^+(v)$ + C $\rightarrow$ AC$(v''')$ + B$^+$ 
\end{itemize}

is given by 

\begin{equation}
\label{sigma}
\sigma_{\text{q,r,d}}(E_k)=2 \pi \int_{0}^{b_{\text{max}}^{q,r,d}(E_k)}P_{\text{q,r,d}}(b,E_k)bdb,
\end{equation}

\noindent
where $b_{\text{max}}^{q,r,d}(E_k)$ indicates the maximum impact parameter for trajectories leading to the process at hand and $P_{\text{q,r,d}}(b,E_k)$ is the opacity function. The opacity function represents the probability of a given process ( $q$, $r$ or $d$) as a function of the impact parameter $b$ and collision energy as

\begin{eqnarray}
P_{\text{q,r,d}}(b,E_k)&=&\int P_{\text{q,r,d}}(b,E_k,\theta,\phi,\eta,\xi)d\Omega \nonumber \\
d\Omega&=&\sin{\theta}d\theta d\phi d\eta d\xi,
\end{eqnarray}

\noindent
which is evaluated through Monte Carlo sampling leading to

\begin{equation}
P_{\text{q,r,d}}(b,E_k)=\frac{N_{\text{q,r,d}}(b,E_k)}{N} \pm \delta_{\text{q,r}}(b,E_k).
\end{equation}

\noindent
Here $N_{\text{q,r,d}}(b,E_k)$ denotes the number of trajectories associated with the process of interest and $N$ stands for the total number of trajectories launched for a given $b$ and $E_k$. Next, applying the one standard deviation rule one finds

\begin{equation}
\label{delta}
\delta_{\text{q,r,d}}(b,E_k)=\frac{N_{\text{q,r,d}}(b,E_k)}{N}\sqrt{\frac{N-N_{\text{q,r,d}}(b,E_k)}{N}}.
\end{equation}

\section{Results}

We run batches of 10$^4$ trajectories per collision energy covering 100 values of the impact parameter, i.e., $N=100$. The trajectories were propagated by solving Eqs~.(\ref{Hamilton1}) and (\ref{Hamilton2}) through the Cash-Karp Runge-Kutta method~\cite{Numericalrecipies} leading to the conservation of the total energy to at least four significant digits while the total angular momentum, $J=|\vec{\rho}_1 \times \vec{P}_1 +\vec{\rho}_2 \times \vec{P}_2 |$ is conserved to at least six digits. In our approach, we assume that the three-body potential energy surface can be described through pair-wise additive potentials as $V(\vec{r}_1,\vec{r}_2,\vec{r}_3)~=~V(\vec{r}_{12})~+~V(\vec{r}_{13})~+~V(\vec{r}_{23})$ (see Fig.~\ref{fig1}). Here we focus on the physical scenario described in Refs.~\cite{Harter2014,Krukow2016}, in which the Rb atoms are spin polarized. Therefore they interact through their triplet potential which is taken from Ref.~\cite{Krych2011}. For the  Ba$^+$-Rb interaction it is assumed that the charged is localized in the Ba atom and the interatomic interaction is described by means of the generalized Lenard-Jones potential: $V(r)=-C_4/r^4(1-1/2(r_m/r^4)^4)$, where $r$ is the atom-ion distance\footnote{For the initial state $r=r_{12}$ (see Fig.~\ref{fig1})}, $C_4$= 160 a.u. and $r_m$=9.27 a$_0$\footnote{This scenario corresponds with the triplet potential energy curve for BaRb$^+$ system, whereas assuming that the charge is localized in the rubidium atom will lead to a singlet ground state potential of BaRb$^+$}. The vibrational states closer to dissociation for BaRb$^+$ based on our model potential are shown in Table~\ref{tab1}. With our model we expect to reproduce properly the progression and density of vibrational states, which is the relevant magnitude for the study of vibrational quenching, since we use the proper physical long-range interaction, however the number and energy of bound states are not the physical ones.  

\begin{table}[h]
\caption{Vibrational bound states for BaRb$^+$ (in mK) assuming the generalized Lenard-Jones potential described in the text.}
\begin{center}
\begin{tabular}{c c}
\hline
$v$&Binding energy ($E_{v}$)  \\ \hline
198 & 0.01 \\
197 & 0.11 \\
196 & 0.43 \\
195 & 1.13 \\
194 & 2.48 \\
193 & 4.77 \\
192 & 8.37 \\
191& 13.70 \\
190 & 21.24 \\
189 & 31.52 \\
188 & 45.15 \\
187 & 62.84 \\
186 & 85.12 \\
185 & 113.03 \\
\hline
\hline
\end{tabular}
\end{center}
\label{tab1}
\end{table}%

In Fig.~\ref{fig2} some trajectories associated with the collision BaRb$^+$(v)~+~Rb are shown. Here one notices that the trajectory shown in panel b leads to a dissociation process, i.e, the molecular bond is broken after the collision. Whereas, the trajectories of panels a and c correspond to quenching events, i.e., the molecular ion in the reactants and products states is the same but in a different vibrational state. However, we would like to stress that the trajectory of panel c shows a more complex situation: after an initial three-body event the two neutrals and the ion form a trimer during 10 ns that finally decays into a BaRb$^+$(v')~+~Rb, resembling the so-called roaming resonances in chemical reactions~\cite{Bowman}. This lifetime is three orders of magnitude larger than the lifetime of He-A complex (where A is a large organic molecule)~\cite{Li2012}. This may be qualitatively understood using the results of Li and Heller~\cite{Li2012}, since these authors found that the lower the collision energy the larger the lifetime of the He-A complex is. Also, it is worth pointing out that the exchange reaction shown in panel (c) of the figure is counted as an inelastic event rather than a reactive one owing the quantal indistinguishability of the atoms. 


Along this section we present the vibrational quenching and dissociation cross section results for RbBa$^+(v)$~+~Rb with $v=187-195$, whose binding energies ranges from 62.84 mK to 1.13 mK based on the model potential employed for BaRb$^+$. This scenario is supported by our previous work on Ba$^+$~- ~Rb~-~Rb three-body recombination~\cite{Krukow2016}, where it was experimentally confirmed that three-body recombination is a very efficient process for molecular ion formation at cold temperatures. Moreover, we found that the internal energy of the molecule correlates with the collision energy between the atomic partners, therefore the vibrational binding energy of the molecular ions should be similar to the typical temperature in cold experiments. Finally we would like to stress that for BaRb$^+$+Rb collisions 17 partial waves are relevant for the dynamics at 1 mK, therefore a QCT calculations is an appropriate method to treat the collision at hand for $E_{\text{k}}\gtrsim$ 1mK.

\begin{figure}[h]
\centering\includegraphics[width=0.8\columnwidth]{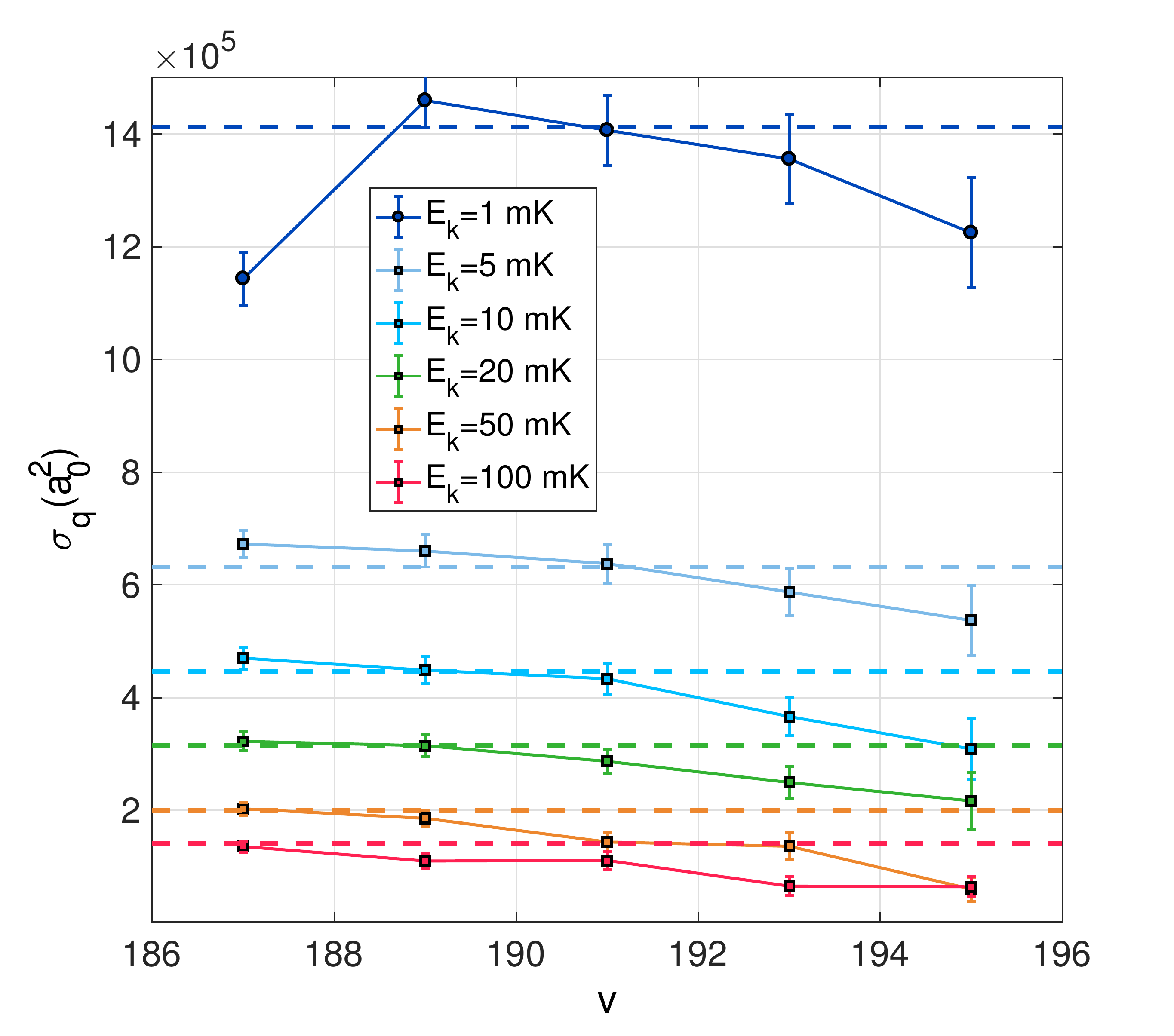}
\caption{Quenching cross section for the collision BaRb$^+(v)$ + Rb $\rightarrow$ BaRb$^+(v' \ne v)$ + Rb as a function of the different initial vibrational states (horizontal axis) and  different collision energies indicated in the figure as the different colors. Dashed lines represent the Langevin cross section for a given energy indicated by the colors in the upper part of of the figure . For this calculations we took $j=0$. The error bars are related with the one standard deviation rule as introduced in Eq.~(\ref{delta}). }
\label{fig3}
\end{figure}

\subsection{Vibrational quenching cross section}
 
 The quenching cross section for RbBa$^+(v)$~+~Rb as a function of the initial vibrational state ($v=187-195$) and different collision energies is shown in Fig.~\ref{fig3}. These results have been obtained through numerical integration Eq.~(\ref{sigma}) by means of the opacity function determined through the Monte Carlo sampling method. Here, one notices that the vibrational quenching cross section for a given collision energy is nearly independent of the initial vibrational state. It also depends on the collision energy as $E_k^{-1/2}$ in agreement with the prediction of the Langevin capture model $\sigma_L(E_k)=\pi(4C_4/E_k)^{1/2}$ (colored dashed lines in Fig.~\ref{fig3}). However, some systematic small deviation between QCT predictions and the Langevin cross section for $v=195$ is observed, as well as for $v=193$ but for $E_k\ge$10~mK.

\begin{figure}[h]
\centering\includegraphics[width=1.0\columnwidth]{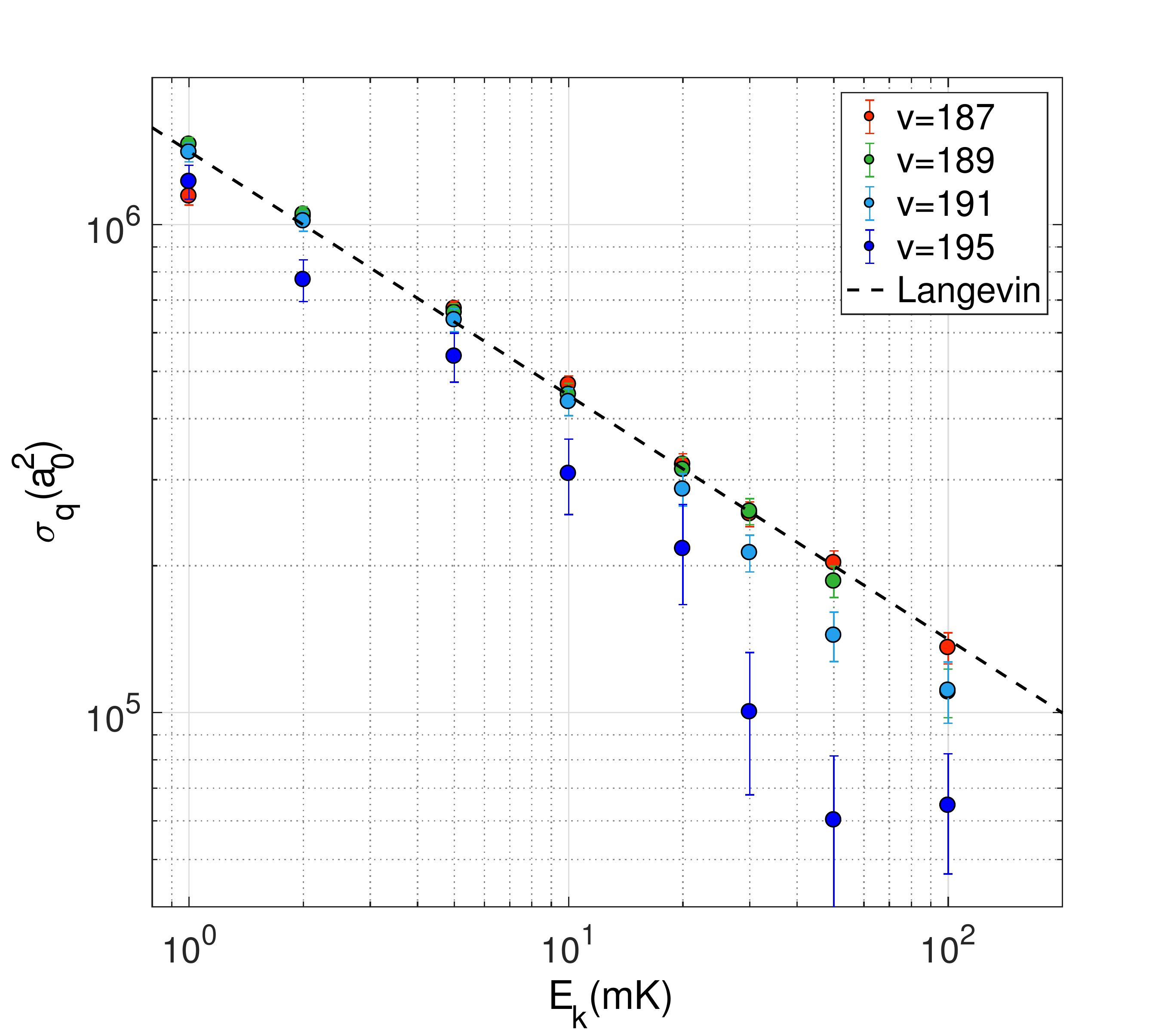}
\caption{Quenching cross section for BaRb$^+(v)$ + Rb $\rightarrow$ BaRb$^+(v' \ne v)$ + Rb as a function of the collision energy $E_k$. The black-dashed lines represent the Langevin cross section. For this calculations we took $j=0$. The error bars are related with the one standard deviation rule as introduced in Eq.~(\ref{delta}). }
\label{fig4}
\end{figure}

In Fig.~\ref{fig4} we show the QCT results for quenching cross section for RbBa$^+(v)$~+~Rb as a function of the collision energy. In this figure we observe that the QCT results for deeply bound initial molecular states agrees with the Langevin prediction, whereas for shallow vibrational states some discrepancies appear. Indeed, these deviations occur at different energies; in particular, shallower vibrational states start to deviate at lower collision energies than more deeply bound vibrational states. These deviations are due to the presence of a new product state channel: dissociation, since the collision energy is enough to break the molecular bound producing three free atoms. This scenario is common to the results shown in Fig.~\ref{fig3}.

\subsection{Dissociation cross section}

\begin{figure}[h]
\centering\includegraphics[width=1.0\columnwidth]{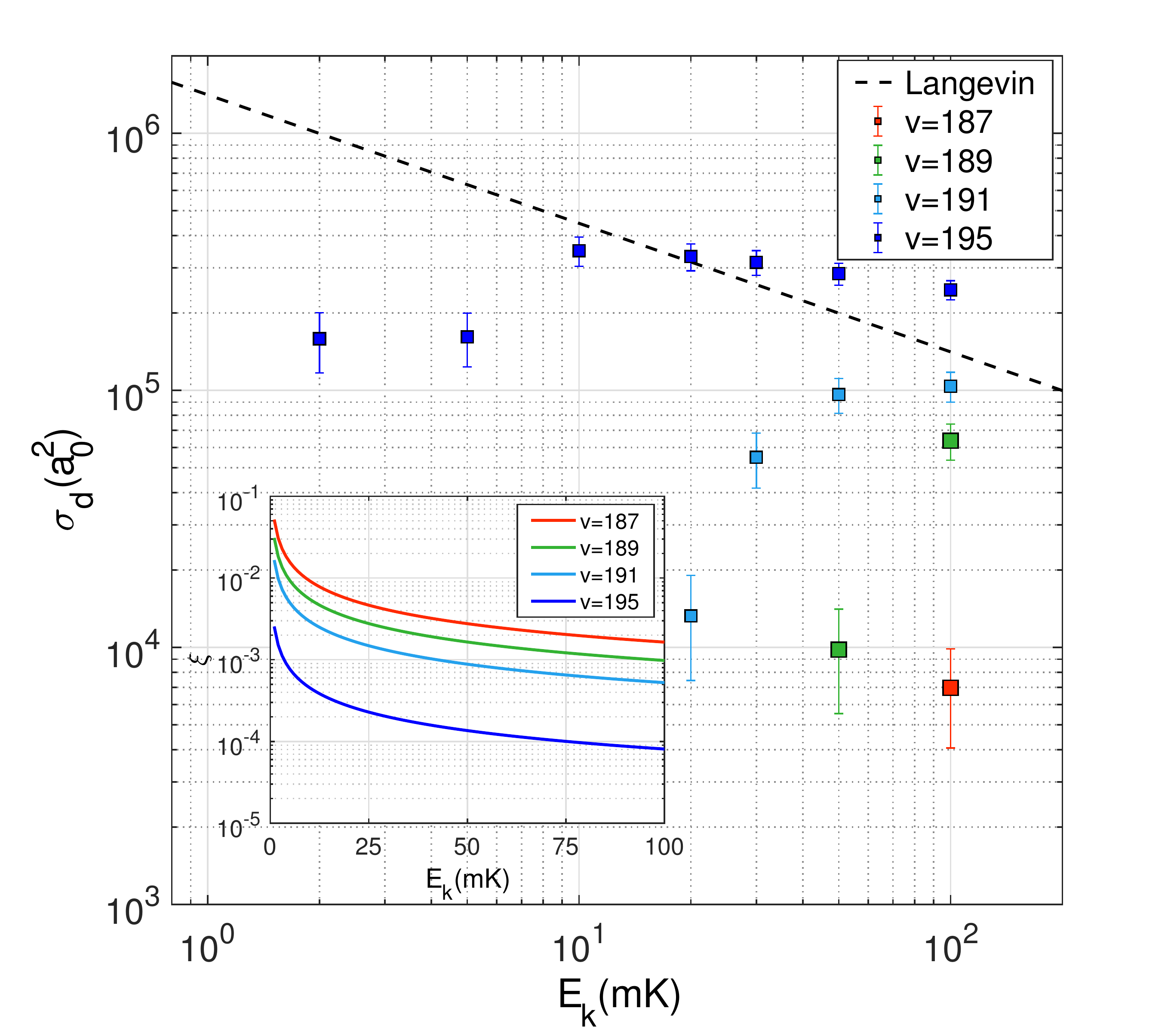}
\caption{Dissociation cross section for the collision BaRb$^+(v)$ + Rb $\rightarrow$ Ba$^+$ + Rb + Rb as a function of the collision energy $E_k$. The different initial vibrational states $v$ are denoted by the different colors. The black-dashed line represents the Langevin cross section. For this calculations we took $j=0$. The error bars are related with the one standard deviation rule as introduced in Eq.~(\ref{delta}). In the inset the adiabaticity parameter as a function of the collision energy is shown. The different colors are related with different initial vibrational states as indicated in the legend.}
\label{fig5}
\end{figure}

In the present case, dissociation is given by the chemical reaction BaRb$^+(v)$ + Rb $\rightarrow$ Ba$^+$ + Rb + Rb, i.e., the colliding Rb atom breaks the molecular bond of the molecular ion. This channel only opens for $E_k> E_{v}$. The dissociation cross section as a function of the collision energy is shown in Fig.~\ref{fig5}, where we see that only for $E_k>E_{v}$ (see table~\ref{tab1}) this channel becomes relevant for the dynamics, as anticipated. Also, we observe that the higher the collision energy the larger the cross section, reaching the Langevin prediction. This behavior is expected since the probability of breaking the molecular bond increases with the collision energy. 


To further understand the efficiency of the energy transfer between translational and internal degrees of freedom we introduce the adiabatic parameter $\xi$~\cite{Levine}, which gives the efficiency of a given energy transfer process by comparing the relevant time scales. In the case at hand, $\xi=\tau_c/\tau_{v,j}$, where $\tau_c$ is the collision time and $\tau_{v,j}$ is the vibrational period introduced in Section~\ref{theory}. Large values of $\xi$ imply that the molecule vibrates many times before the collision happens indicating an strong spring constant and hence an inefficient transfer. For $\xi\approx1$ the energy transfer becomes efficient due to the synchronization between collision and vibration, but it is only when $\xi\ll1$ that the molecule shows a negligible vibrational motion during the collision and approaches to the limit of a weak oscillator. This leads to a high efficiency energy transfer~\cite{Levine}. In our scenario, $\xi\ll1$ or the weak oscillator spring limit will translate into efficient quenching and dissociation. 

In the inset of Fig.~\ref{fig5} $\xi$ as a function of the collision energy is shown. For these results we take $\tau_c=b_L/v_{k}$, where $b_L=(2\alpha/E_k)^{1/4}$ is the Langevin impact parameter and $v_k$ represents the collision velocity. In this figure, we observe that $\xi\ll1$ for all the vibrational states and the whole range of collision energies, thus vibrational quenching and dissociation processes should be efficient as was revealed by the QCT calculations in Figs.~\ref{fig3}, \ref{fig4}~- \ref{fig5}. In the particular case of dissociation, apart from being efficient it needs to be energetically available, as the results of Fig.~ \ref{fig5} show.

\begin{figure}[h]
\centering\includegraphics[width=1.0\columnwidth]{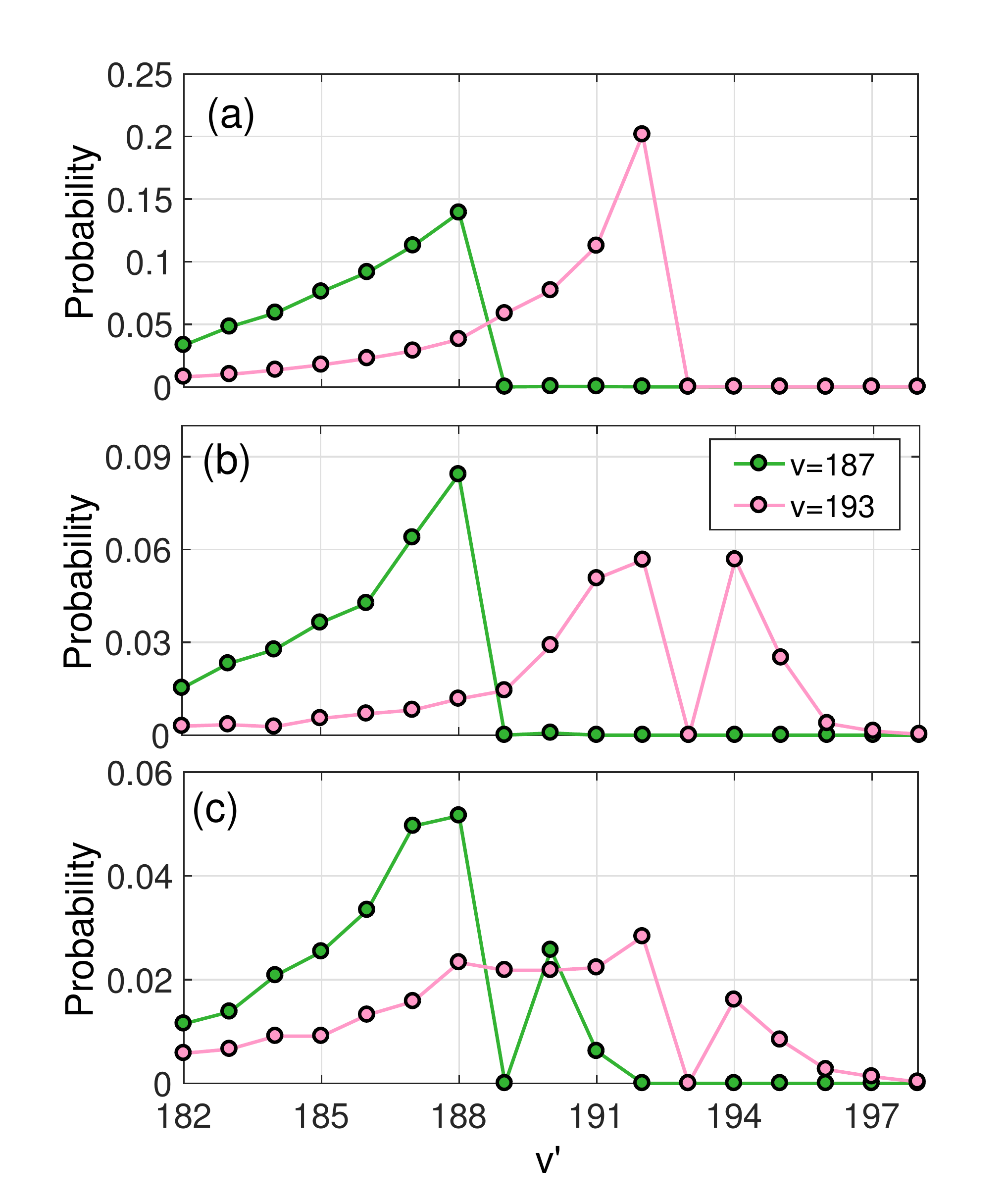}
\caption{Vibrational and rotational distribution for the product states after the quenching process  BaRb$^+(v)$ + Rb $\rightarrow$ BaRb$^+(v' \ne v)$ + Rb for different energies: panel (a) $E_k$=1mK,  panels (b) $E_k$=10mK and panels (c) $E_k$=10mK.}
\label{fig6}
\end{figure}

\subsection{Distribution of final states}

The vibrational quenching cross section accounts for all the different trajectories leading to the molecular state with a different vibrational state, however it does not provide information about the state-to-state processes. In other words, it does not say anything about the energy distribution of the product states. From the opacity function, we have calculated the probability of finding the molecular ion in a given $v'$ state (BaRb$^+(v)$ + Rb $\rightarrow$ BaRb$^+(v' \ne v)$ + Rb), and the results are shown in Fig.~\ref{fig6} for different collision energies. In this figure one notices a broad vibrational distribution of the final states, although a single quanta vibrational de-excitation is the most probable final state after a quenching process. In particular, in panel (a) at $E_k$=1~mK there is not an available excited vibrational state. In panel (b) at $E_k$=10~mK, some excited vibrational states for $v=193$ become available and they show a decent probability of 7$\%$ owing the high efficiency for vibrational energy transfer; the same applies in panel (c) with $E_k$=20~mK, but now some excited vibrational states are also available for $v=189$. Also, it is worth observing that as the collision energy grows the overall amplitude of the vibrational distribution decreases due to the dissociation channel. 


\subsection{A reactive channel:Rb$_2$ as the final product state}

\begin{figure}[h]
\centering\includegraphics[width=1.0\columnwidth]{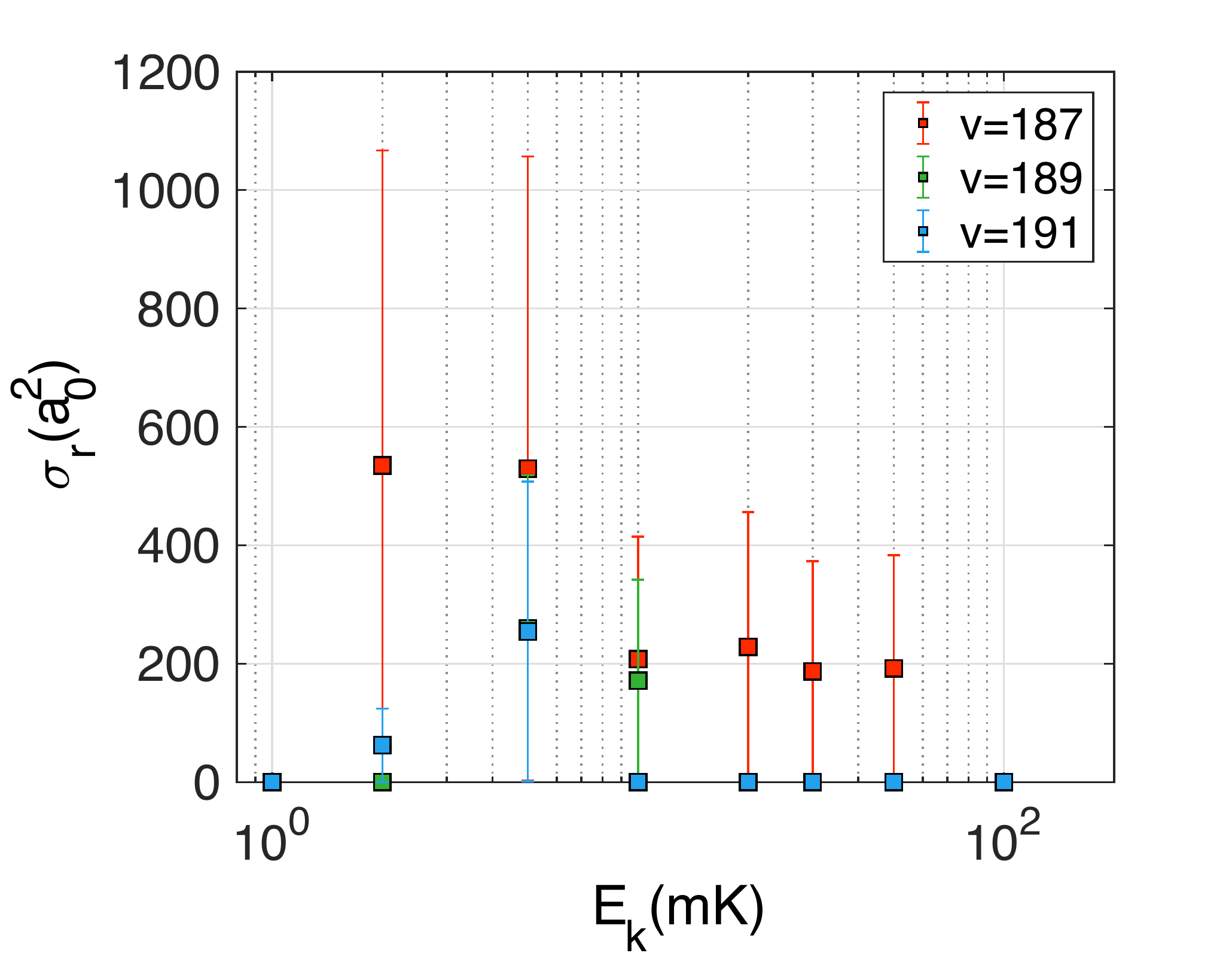}
\caption{Vibrational and rotational distribution for the product states after the quenching process  BaRb$^+(v)$ + Rb $\rightarrow$ BaRb$^+(v' \ne v)$ + Rb for different energies: panel (a) $E_k$=1mK,  panels (b) $E_k$=10mK and panels (c) $E_k$=10mK.}
\label{fig7}
\end{figure}

In our previous work on ion-atom-atom three-body recombination a threshold law for the three-body recombination rate was predicted. This implied that after a three-body event molecular ion association is more preferable than neutral molecule formation~\cite{JPR2015}. The prediction was experimentally confirmed by Kr\"ukow et al.~\cite{Krukow2016}. Here, the reactive process BaRb$^+(v)$ + Rb $\rightarrow$ Rb$_2(v'')$ + Ba$^+$ is a possible reaction pathway which has been considered and the results are shown in Fig.~\ref{fig7}.

In Fig.~\ref{fig7} it is noticed that the formation of Rb$_2$ is $\sim$ 10$^{3}$ times less probable than vibrational quenching or dissociation (see Figs.\ref{fig4} and \ref{fig5}, which confirm that the ion-neutral interaction is the dominant interaction of the system as was previously demonstrated for three-body recombination~\cite{JPR2015,Krukow2016}. Therefore, we can conclude that in ion-atom-atom systems the ion prefers to form a molecule with a neutral than to be free after an molecular ion-atom collision. Nevertheless, from the figure it is interesting to point out that for loosely bound vibrational states of the molecular ion, the formation of Rb$_2$ seems to be negligible whereas deeply bound states exhibit the opposite behavior. Certainly this is an interesting trend that should be addressed in a future work. 



\section{Sympathetic cooling of molecular ions}

\begin{figure}[h]
\centering\includegraphics[width=1.0\columnwidth]{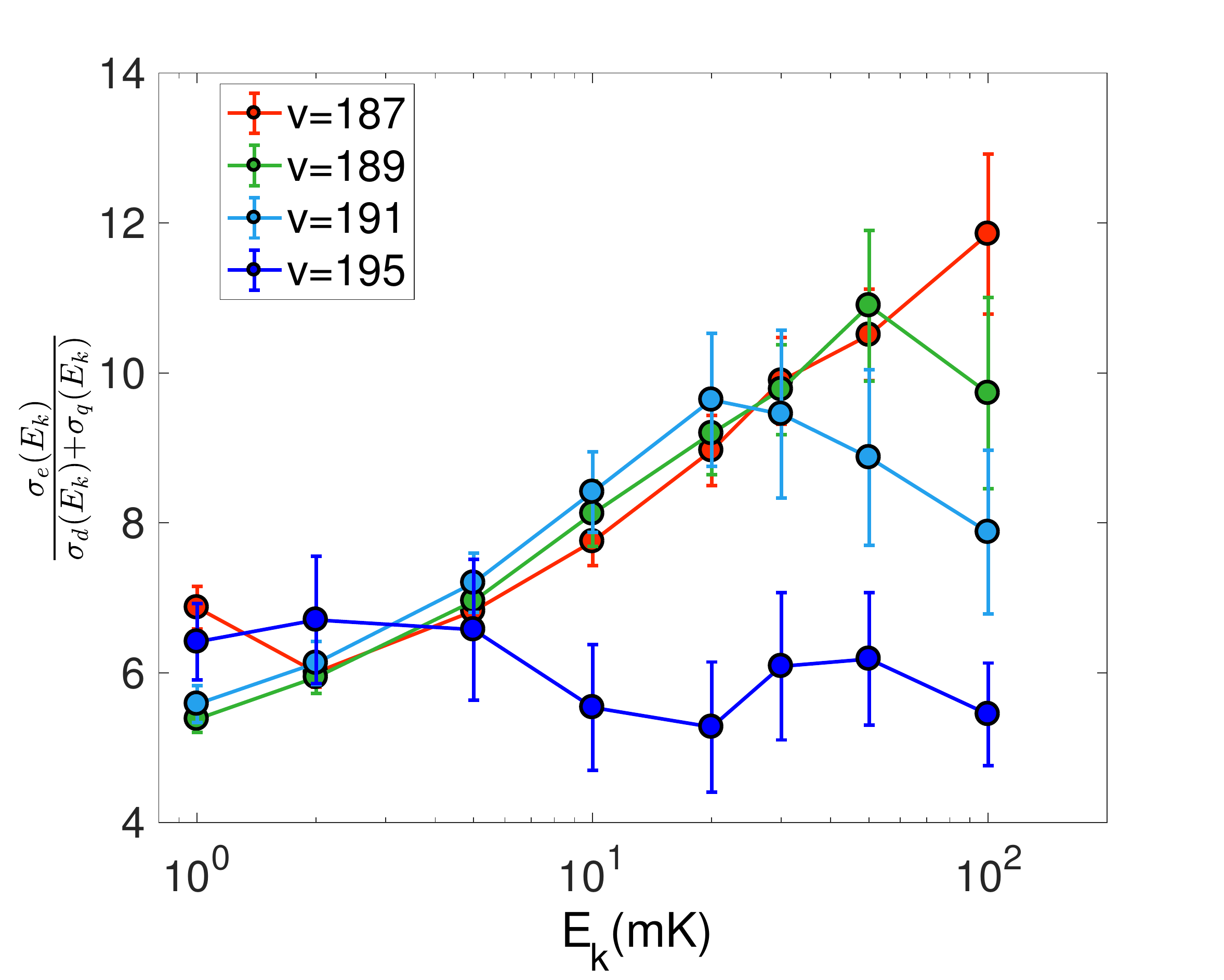}
\caption{Elastic to inelastic ration for BaRb$^+(v)$ + Rb as a function of the different initial vibrational states (see legend). For this calculations $j=0$. The error bars are related with the one standard deviation rule as introduced in Eq.~(\ref{delta}). }
\label{fig8}
\end{figure}

Sympathetic cooling becomes operative when the elastic to inelastic/reactive collision is large. In other words, when elastic collisions (thermalizing collisions) are more frequent than inelastic processes involving exothermic and endothermic pathways. By means of QCT calculations we have calculated the elastic-to-inelastic ratio (including inelastic and reactive processes) for BaRb$^+(v)$~+~Rb as a function of different initial vibrational states for wide range of collision energies; the results are shown in Fig.~\ref{fig8}. Here the elastic cross section is given by 

\begin{equation}
\sigma_e(E_k)=\pi \left(\frac{\mu \alpha^2}{\hbar^2} \right)^{1/2}\left(1+\frac{\pi^2}{16} \right)E_k^{-1/3},
\end{equation}

\noindent
following the semiclassical approach of C\^{o}t\'e and Dalgarno~\cite{Cote2000}, whereas the inelastic cross section has been taken as the quenching cross section plus the dissociation cross section.

Fig.~\ref{fig8} shows that the elastic-to-inelastic ratio grows with the collision energy until it reaches a maximum and then it drops, with the exception of $v$=195, where a nearly flat behavior is observed. The overall growth is due to the different energy scaling between the elastic ($\propto E_k^{-1/3}$) and inelastic cross sections ($\propto E_k^{-1/2}$), leading to a elastic-to-inelastic ratio $\propto E_k^{1/6}$) which corresponds to the trend observed in the figure. The maximum in the magnitude at hand occurs when the dissociation channel becomes energetically available (see Fig.~\ref{fig5}) and the two inelastic channels become operative, leading to an enhancement of the loss mechanisms. For $v=$195 the elastic-to-inelastic ratio shows a flat behavior since in the whole collision energy range covered in this figure (with exception of the first)  dissociation and quenching are readily available leading to a smaller and almost flat elastic-to-inelastic ratio. This behavior is pretty different from the results that Stoecklin et al.~\cite{Hudson2016} have reported on vibrational quenching for deeply bound vibrational states. In particular, the authors found that the elastic-to-inelastic ratio increases as the collision energy gets smaller. This discrepancy may be due to the fact that for loosely bound vibrational states the efficiency of the quenching is given by the Langevin capture model, whereas for deeply bound vibrational states some deviations occur due to the role of anisotropy in the atom-molecule interaction. 

Assuming a head-on collision of a "hot" molecular ion with an ultracold atom (assumed at rest), the molecule will transfer a certain amount of its energy to the atom given by

\begin{equation}
\label{ratio}
\frac{E_T}{E_{in}}=\frac{4m_{\text{BaRb}^+}m_{\text{Rb}}}{(m_{\text{Rb}}+m_{\text{BaRb}^+})^2},
\end{equation}

\noindent
with $m_{\text{Rb}}$ is the mass of rubidium, $m_{\text{BaRb}^+}$ is the mass of the molecular ion, $E_{in}$ is the kinetic energy of the molecule in the lab frame and $E_T$ is the kinetic energy of the atom after the collision in the lab frame. Then, the final energy of the molecular ion after $N$ elastic collisions will be $(1-E_T/E_{in})^N$ Following the results in Fig.~\ref{fig8}, the elastic-to-inelastic ratio is $\sim 5-10$, i.e., 5 to 10 elastic collision per inelastic event, thus taking $N=5$ the final energy of the molecular ion will be 3.2$\times$ 10$^-4$ $E_{in}$, which represents an excellent transfer efficiency. However, in a cold chemistry experiment the ions are held in a Paul trap where micromotion is a continuous energy source on the system, that certainly will affect the capabilities of the molecular ion to cool down. However, the study of this particular matter is beyond the scope of the present work. 

\section{Conclusions}

We have applied the celebrated QCT calculation method to the study of vibrational quenching and dissociation of molecular ions in loosely bound vibrational states colliding with neutral atoms at cold temperatures. In particular we have focussed on BaRb$^+(v)$ + Rb due to its experimental relevance in ion-neutral-neutral three-body recombination experiments at cold temperatures. The validity of a quasi-classical approach has been shown for collision energies $\gtrsim$ 1mK owing the 17 partial waves that contribute to the cross section. As a result, we have shown that QCT calculations-based vibrational quenching cross section follows the Langevin capture model independently of the vibrational state of the molecular ion, this translates into a very efficient energy transfer between translational and vibrational degrees of freedom. 

Our study complements previous studies employing coupled-channel method for vibrational quenching of molecular ions in deeply bound vibrational states colliding with ultracold atoms~\cite{Hudson2016,Gianturco2011,Wester2015,Stoecklin2005,Stoecklin2008,Stoecklin2011}. Indeed, employing a coupled-channel approach here would be extremely computationally heavy due to large size of the Hilbert space, as well as the large number of partial waves must be included. Thus, QCT seems to be a very good and reliable method for studying collisional processes of vibrationally excited molecular ions colliding with ultracold atoms. Finally, our results have implications in experiments for ion-neutral-neutral three-body recombination to assist in the understanding of the role of the trapping laser in the dynamics of the molecular ion~\cite{Krukow2016}, as well as its inherent importance in understanding the fundamentals of molecular ion-neutral collisional processes. 

\section{Acknowledgements}

We would like to acknowledge Amir Mohammadi and Johannes Denschlag for motivating this work through fruitful discussions and elucidating the experimental implications of our work, as well as Matthew T.  Eiles and Francis Robicheaux for carefully reading the manuscript prior to publication. 

\bibliography{theory}

\end{document}